 \documentclass[preprint,floats,aps,epsfig,nofootinbib,amssymb]{revtex4}
\usepackage{mathrsfs}

\usepackage{slashed}
\usepackage{graphicx,color}
\usepackage{dcolumn}
\usepackage{bm}
\usepackage{subfig}
\usepackage{graphicx}
\usepackage{amssymb}

\def\lsim{\mathrel{\rlap{\lower4pt\hbox{\hskip1pt$\sim$}}
    \raise1pt\hbox{$<$}}}         %less than or approx. symbol
\def\gsim{\mathrel{\rlap{\lower4pt\hbox{\hskip1pt$\sim$}}
    \raise1pt\hbox{$>$}}}         %greater than or approx. symbol

\begin{document}

\vspace*{-5.8ex}
\hspace*{\fill}{ACFI-T14-21}

\vspace*{+3.8ex}

\title{Electroweak Baryogenesis in the Exceptional Supersymmetric Standard Model}

\author{Wei Chao$^{}$}
\email{chao@physics.umass.edu}

 \affiliation{ Amherst Center for Fundamental Interactions, Department of Physics, University of Massachusetts-Amherst
Amherst, MA 01003 }

\vspace{3cm}

\begin{abstract}

We study electroweak baryogenesis in the $E_6$ inspired exceptional supersymmetric standard model (${\rm E_6SSM}$).  The relaxation coefficients driven by singlinos and the new gaugino as well as the transport equation of the Higgs supermultiplet number density in the ${\rm  E_6SSM}$ are calculated. Our numerical simulation shows that both CP-violating source terms from singlinos and the new gaugino can solely give rise to a correct baryon asymmetry of the Universe  via the electroweak baryogenesis mechanism.

\end{abstract}

\maketitle
\section{Introduction}

The origin of the baryon asymmetry of the Universe (BAU) is one of the longstanding problems in particle physics and cosmology. Combining the WMAP seven year results~\cite{Komatsu:2010fb} with those from  CMB and large scale structure measurements one has
\begin{eqnarray}
Y_B\equiv{\rho_B \over s } = (8.82\pm 0.23) \times 10^{-11} \label{expp}
\end{eqnarray}
where $\rho_B$ is the baryon number density, $s$ is the entropy density of the Universe. The recent results obtained by the Planck satellite are consistent, giving $Y_B = (8.59 \pm 0.11) \times 10^{-11}$~\cite{Ade:2013zuv}. 

Assuming that the Universe was matter-antimatter symmetric at its birth, it is reasonable to suppose that interactions involving elementary particles generated the BAU during the subsequent cosmological evolution. To generate the observed BAU, three Sakharov criteria~\cite{Sakharov:1967dj} must be satisfied in the early Universe: (1) baryon number violation; (2) C and CP violation; (3) a departure from the thermal equilibrium (assuming exact CPT invariance). These requirements are realizable, though doing so requires physics beyond the Standard Model (SM). To that end, theorists have proposed a variety of baryogenesis scenarios whose realization spans the breadth of cosmic history. Electroweak baryogenesis (EWBG)~\cite{Morrissey:2012db,Cohen:1993nk,Trodden:1998ym,Riotto:1998bt,Riotto:1999yt,Quiros:1999jp,Dine:2003ax,Cline:2006ts}  is one of the most attractive and promising such scenarios, and it is generally the most testable with a combination of searches for new degrees of freedom at the LHC and low-energy tests of CP invariance.

Successful EWBG requires a first order electroweak phase transition and sufficiently effective CP violation during the transition. Neither requirement is satisfied in the SM. One simple extension of the SM that may allow them to be satisfied is the two Higgs doublet model (2HDM) (for a recent review, see Ref.~\cite{Branco:2011iw}). The minimal supersymmetric standard model (MSSM), as a typical type-II 2HDM, is an attractive framework for EWBG and other new physics, since it can provide elegant explanations for questions that can not be accommodated in the SM. In the MSSM, the light stop scenario is necessary for successful baryogenesis. But the Higgs mass  discovered by the CERN LHC~\cite{atlas,Aad:2013wqa,cms,Chatrchyan:2013lba} as well as Higgs decay rates put strong constraint on the parameter space of the light stop, which allows one to exclude the EWBG in the MSSM at  quite high confidence level~\cite{Curtin:2012aa}\footnote{For more precise statements of the EWBG in the MSSM, see Ref.~\cite{Curtin:2012aa} for detail. }.

Possible ways out are considering extensions of the MSSM, of which the Next to Minimal Supersymmetric Standard Model (NMSSM)~\cite{Ellwanger:2009dp} is attractive since it provides solution to the $\mu $ problem and naturally accommodates a $125~{\rm GeV}$ SM Higgs. Constraints on the parameter space of the model from the non-observation of permanent  electric dipole moments (EDMs) of neutron, Mercury, Thallium,   deuteron and Radium as well as the parameter space available for singlino driven electroweak baryogenesis were studied in Ref.~\cite{Cheung:2011wn,Cheung:2012pg}. In this paper we study the electroweak baryogenesis in the exceptional superysymmetric extension to the SM~\cite{King:2005jy}, ${\rm E_6SSM}$, which is a string theory inspired supersymmetric model based on an $E_6$ grand unification (GUT) group.  ${\rm E_6SSM}$ also accommodates a $125$~GeV Higgs \cite{Athron:2013ipa} and can dynamically generate the $\mu$ term.  The Higgs sector of the ${\rm E_6SSM}$  contains $9$ Higgs supermultiplets. As a result, it can naturally  derive a strongly first order EWPT.  The chargino sector in  the ${\rm E_6SSM}$, although  extended compared with the MSSM case, gives no extra contribution the CP-violating source term as required by the EWBG mechanism.  The neutralino sector in the ${\rm E_6SSM}$ is greatly enlarged compared with the MSSM and NMSSM cases. We calculate in the section III the relaxation coefficients driven by singlinos and the new gaugino in the neutralino sector. The transport equation of the Higgs supermultiplet number density is solved using analytical approximation. Our numerical simulation shows that  CP violation sources from the new gaugino and singlinos in the neutralino sector may solely give rise to a baryon asymmetric Universe via the EWBG, without conflicting with the EDM constraints.

 The paper is organized as follows: In section II we give a brief introduction to the ${\rm E_6SSM}$. Section III is devoted to the investigation of EWBG induced by the neutralino sector of the ${\rm E_6SSM}$. We summarize in section IV.
 
 \section{${\rm E_6SSM}$}
 
 The ${\rm E_6SSM}$, which originates from an $E_6$ GUT or string theory in extra dimensions, involves a unique choice for the extra Abelian gauge group namely $U(1)_N$.  To ensure anomaly cancellation, the particle content of the ${\rm E_6SSM}$ includes three complete fundamental $27$ representations of $E_6$, which decompose under the $SU(5)\times U(1)_N$ subgroup  as follows. 
 \begin{eqnarray}
 27 \to \left(10, {1\over \sqrt{40 }}\right)\oplus \left(\bar 5 , {2 \over \sqrt{40 }}\right) \oplus \left(\bar 5 , -{3 \over \sqrt{40 }}\right) \oplus \left(\bar 5 , -{2 \over \sqrt{40 }}\right) \oplus  \left(1  , {5 \over \sqrt{40 }}\right) \oplus \left(1, {\over }0 \right) \label{decompose}
 \end{eqnarray}
 where the first and second quantities in the brackets are the $SU(5)$ representation and $U(1)_N$ charge respectively. The first two terms on the right side of Eq. (\ref{decompose}) contain all the matter contents, the third and fourth terms contain the pair of  Higgs doublets as well as diquarks with electric charges $-1/3$ ad $+1/3$, respectively. Scalar singlet, the third generation of which breaks the $U(1)_N$ gauge symmetry spontaneously, is contained in the fifth term. Right handed neutrinos is associated with the last term.

 In $E_6$ models the renormalizable superpotential comes from $27\times 27 \times 27$ decomposition of the $E_6$ fundamental representation. The superpotential,  that respects to $SU(3)_C\times SU(2)_L\times U(1)_Y$, can be written as~\cite{King:2005jy,King:2005my}
 \begin{eqnarray}
 W_0 &=& \lambda_{ijk} S_i H^u_j H^d_k + \kappa_{ijk} S_i D_j D_k +Y_{ijk}^N N_i H^u_j L_k \nonumber \\
 &+& Y^u_{ijk} u_i H^u_j  Q_k +Y^d_{ijk} d_i H^d_j  Q_k  + Y^e_{ijk} E_i H^d_j  L_k  \\
 W_1 &=& g_{ijk}^u D_i Q_j Q_k + g^d_{ijk} D_i d_j u_k  \label{diquark}
 \end{eqnarray}
where we have assumed that exotic quark $D_i$ are diquarks, which carry a twice large baryon number than the ordinary quark fields, and can decay into the SM quarks via interactions in (\ref{diquark}).  $D_i$ and $\bar D_i $ can  also be  leptoquarks, in which case  superpotential in (\ref{diquark})  will be replaced with  Yukawa interactions between  $D_i$, lepton and quark superfields. 

In the ${\rm E_6SSM}$ the neutralino sector is extended to include eight additional neutral components: $\tilde S^i $, $\tilde H^\alpha_u $, $\tilde H_d^\alpha $ and $\tilde B^\prime $, where $i=1,2,3$ and $\alpha =1,2$. The neutralino mass matrix can be written as~\cite{Hall:2011yb} 
\begin{eqnarray}
M^{\rm N}_{\rm E_6SSM}=\left( \matrix{ M^{\rm N }_{\rm USSM} &B_2 & B_1 \cr  B_2^T & A_{22} & A_{21} \cr B_1^T & A_{21}^T & A_{11} }\right)
\end{eqnarray}
in the basis
\begin{eqnarray}
\left(\matrix{ \tilde B & \tilde W^3 & \tilde H^3_d & \tilde H^3 _u & \tilde S& \tilde B^\prime & |&\tilde H^2_d & \tilde H_u^2 & \tilde S^2& |&\tilde H^1_d & \tilde H_u^1 & \tilde S_1 }\right).
\end{eqnarray}
The submatrix $M_{\rm USSM}^{\rm N}$~\cite{Kalinowski:2008iq}  is the neutralino mass matrix in the USSM. Expressions of other submatrices can be found in \cite{Hall:2011yb}.  We refer to \cite{King:2005jy,King:2005my,Hall:2011yb,Kalinowski:2008iq,Athron:2009bs,Athron:2009ue,King:2008qb,Athron:2012sq,Athron:2013ipa}  for studies of low energy phenomenologies of the ${\rm E_6SSM}$. 

\section{Electroweak baryogenesis}

In the EWBG scenario, three Sakharov conditions are realized in the following way~\cite{Morrissey:2012db,Chao:2014dpa,Chung:2009cb}: First, the scalar sector of the ${\rm E_6SSM}$ gives rise to a  strongly first order electroweak phase transition, which provides a departure from thermal equilibrium at temperature $T\sim 100 ~{\rm GeV}$. During the EWPT, bubbles of broken electroweak symmetry nucleate and expand in a background of unbroken symmetry, filling the Universe to complete the phase transition. Second, CP-asymmetry charge density is produced by the CP-violating interactions of the Higgsinos at the walls of the expanding bubbles, where the Higgs vacuum expectation value is space-time dependent, This CP-asymmetry diffuses ahead of the advancing bubble and is converted into a net density of left-handed fermions, through inelastic interaction in the plasma. Third, baryon number is violated by the sphaleron processes. The presence of nonzero $n_L$ biases the sphaleron processes, resulting in the production of the baryon asymmetry.

Ordinary quantum field theory is not appropriate for treating the microscopic dynamics of the electroweak phase transition, since the non-adiabatic evolution of states and the presence of degeneracies in the spectrum break the zero-temperature equilibrium relation between the in- and out- states. We derive the source terms in the quantum transport equation based on the closed time path formulation of non-equilibrium quantum field theory~\cite{Schwinger:1960qe,Chou:1984es}.  The equations governing the space-time dependence of number densities of a given spaces can be written as~\cite{Riotto:1998zb,Lee:2004we,Chung:2009qs,Cirigliano:2006wh}
\begin{eqnarray}
{\partial n \over \partial t } +\bigtriangledown\cdot j(x) &=& -\int d^3 z \int_{- \infty}^{ x_0 } d z^0 {\rm Tr } [ \Sigma^> (x,z) S^< (z,x) -S^> (x,z)\Sigma^< (z,x) \nonumber \\
&&+ S^<(x,z) \Sigma^>(z,x) -\Sigma^<(x,z) S^>(z,x) ] 
\end{eqnarray}
where self energy $\Sigma^{<,>}$ encode all the information about particle interactions.

The rephasing invariant combinations in the ${\rm E_6SSM}$ relevant to electroweak baryogensis can be written as $ \phi_{\lambda_{ijk}} + \phi_{A_{\lambda_{ijk}}}$, $\phi_{M_1} $, $\phi_{M_2}$ and $\phi_{M_1^\prime}$, where $A_{\lambda_{ijk}}$ are couplings of trilinear interactions $S_i H_{uj} H_{d k}$ in the soft supersymmetry breaking lagrangian. $M_1,~M_2$ and $M_1^\prime$ are the mass of bino, wino, and the new gaugino respectively. For simplicity we set  $A_{\lambda_{ijk}}$ to be real in our calculation.
We define the four component spinors as
\begin{eqnarray}
&& \Psi_{\tilde H_i^+ }  =\left(\matrix{ \tilde H_{u_i}^+ \cr \tilde H_{d_i}^{-\dagger }} \right) \; , \hspace{0.5cm} \Psi_{\tilde W^0} =\left( \matrix{\tilde W_3 \cr \tilde W_3^\dagger } \right)\; ,
\hspace{0.5cm} \Psi_{\tilde W^+} = \left( \matrix{ \tilde W^+ \cr \tilde W^{-\dagger}}  \right) \; , \hspace{0.5cm} \Psi_{\tilde B} = \left(  \matrix{\tilde B \cr \tilde B^\dagger }\right) \nonumber  \\
&&\Psi_{\tilde H_i^0} = \left( \matrix{ - \tilde H_{u_i}^0 \cr \tilde H_{d_i}^{0\dagger}}\right) \; , \hspace{0.5cm} \Psi_{\tilde S_i } = \left(  \matrix{\tilde S_i \cr \tilde S_i^\dagger }\right) \; .
\end{eqnarray}
The Higgsino-gaugino-VEV interactions can be written as
\begin{eqnarray}
{ \cal L}^{\rm int} &=& -g_2 \overline{\tilde \Psi_{\tilde H_3^+ }} \left(  v_d P_L +v_u e^{ i \phi } P_R \right) \Psi_{\tilde W^+ }-{1 \over \sqrt{2}} \overline{\Psi_{\tilde H_3^0 }} \left(  v_d P_L + v_u P_R e^{i \phi} \right) ( g_2 \Psi_{\tilde W^0 } - g_1 \Psi_{\tilde B }) 
\nonumber \\
&& + \overline{\Psi_{\tilde H_i^0 }} \left\{ |\lambda_{j3i}|v_u P_L e^{ i\arg{\lambda_{j3i}}}  - |\lambda_{ji3}|v_d P_R e^{ i(-\arg \lambda_{ji3} + \arg(\lambda_{3jj})}\right \} \Psi_{\tilde S_j }  \nonumber \\
&&-\sqrt{2} g_N^{} \overline{\tilde H_3^0} \left(  e^{- i\varphi_{M_1^\prime}/2 } v_d Q_N^d P_L - e^{ i( \phi+\varphi_{M_1^\prime}/2 )} v_u^{} Q_N^u P_R^{}   \right) \Psi_{\tilde{B}^\prime}
\end{eqnarray}
where $\phi =\arg \lambda_{333}$ and $\varphi_{M_1^\prime} =\arg M_1^\prime$, $Q_N^d$ and $Q_N^u$ are charges of $H_{ui}$ and $H_{di}$ under the $U(1)_N$. The first two terms are the same as those in the MSSM, the third term is the interactions between Higgsinos and singlinos, the last term is the interaction of the third generation Higgsino with the new gaugino.

We ignore the wall curvature in our analysis so all relevant functions depend on the variable $\bar z = z +v_w t$, where $v_w$ is the wall velocity; $\bar z<0, >0 $ correspond to the unbroken and broken phases, respectively. Working in the closed time path  formulation and under the ``vev-insertion" approximation~\cite{Riotto:1998zb,Lee:2004we,Chung:2009qs,Cirigliano:2006wh}, we compute the CP-violating source induced by singlinos and the new gaugino mediated processes ($\tilde H \to \tilde S\to \tilde H$ and $\tilde H \to \tilde B^\prime\to \tilde H$),
\begin{eqnarray}
S_{\tilde S }^{\rm CPV}& =&-2 \sum_{ij=1}^{3} |\lambda_{ij3} \lambda_{i3j}| \sin[ \arg(\lambda_{ij3})+ \arg(\lambda_{i3j} ) -\arg (\lambda_{3jj})] v^2 \dot{\beta}  |m_{\tilde S_i }| |m_{\tilde H_j }|\nonumber \\
&\times&\int k^2 dk  {1 \over \pi^2 \omega_{\tilde{s}_i } \omega_{\tilde{H}_j} }  {\rm Im}\left\{ { n(\varepsilon_{\tilde{S}_i}) -n(\varepsilon_{\tilde H_j}^* ) \over (\varepsilon_{\tilde S_i } -\varepsilon_{\tilde H_j }^*)^2 } -{ n(\varepsilon_{\tilde{S}_i}) +n(\varepsilon_{\tilde H_j} ) \over (\varepsilon_{\tilde S_i } +\varepsilon_{\tilde H_j }^{})^2 } \right\}   \label{xxx}\\
S_{\tilde B^\prime }^{\rm CPV}&=&-{3 \over 5}g_1^2\sin[ \arg(\lambda_{333}  m_{\tilde Z^\prime })] v^2 \dot{\beta}  |m_{\tilde Z^\prime }| |m_{\tilde H_3 }| \nonumber  \\
&\times&\int k^2 dk  {1 \over \pi^2 \omega_{\tilde{Z}^\prime_i } \omega_{\tilde{H}_3} } {\rm Im }\left\{ { n(\varepsilon_{\tilde{Z}^\prime}) -n(\varepsilon_{\tilde H_3}^* ) \over (\varepsilon_{\tilde Z^\prime} -\varepsilon_{\tilde H_3}^*)^2 } -{ n(\varepsilon_{\tilde{Z}^\prime}) +n(\varepsilon_{\tilde H_j} ) \over (\varepsilon_{\tilde Z^\prime } +\varepsilon_{\tilde H_3 }^{})^2 } \right\}  \label{xxxxx}
\end{eqnarray}
where $n(x) =1/(\exp(x)+1)$, being the fermion distribution function;  $\varepsilon_{\tilde H ,\tilde S} =\omega_{\tilde H ,\tilde S} -i\Gamma_{\tilde H ,\tilde S}$ are complex poles of the spectral function with $\omega^2_{\tilde H ,\tilde S}={k^2 +m^2_{\tilde H ,\tilde S}}$, where $m_{\tilde H ,\tilde S}$ and $\Gamma_{\tilde H ,\tilde S}$ are the thermal masses and thermal rates of $\tilde H $ and $\tilde S$, respectively.  Before proceeding, we note that the VEV insertion approximation used in obtaining eqs.~(\ref{xxx},\ref{xxxxx}) is likely to lead to an overly large baryon asymmetry by at least a factor of a few, though a definitive quantitative treatment of the CPV fermion sources remains an open problem. The results quoted here, thus, provide a conservative basis for restrictions on the EWBG-viable parameter space. For a detailed discussion of the theoretical issues associated with the computation of the CPV source terms, see Ref.~\cite{Morrissey:2012db} and references therein. The thermal mass of the singlinos and the new gaugino can be written as
\begin{eqnarray}
m_{\tilde S_i }^2 &\approx&\left( { {5 \over 64} g_N^2 + \sum_{jk} {1\over 32} |\lambda_{ijk}|^2 } \right)T^2  \; ,  \\
m_{\tilde Z^\prime}^2  &\approx& M_{\tilde Z^\prime }^{ 2 } + {27 \over 32 } g_N^2 T^2 \; ,
\end{eqnarray}
where $g_N $ is the gauge coupling of the $U(1)_N$. Notice that  CP violating source term  in (\ref{xxx})  is is closely related to  Debye masses of singlinos.  

The CP-conserving terms can be written as $S^{CP}_{\tilde H_j } = \sum_i ( \Gamma_{\tilde H_j } (\mu_{\tilde S_i } + \mu_{\tilde H_j } ) + \Gamma_{\tilde H_j }^-  (\mu_{\tilde S_i } - \mu_{\tilde H_j } )) $, where 
\begin{eqnarray}
\Gamma_{\tilde H_j }^{\pm}&=& {1 \over T } \int {k^2 d k \over 2 \pi^2  \omega_{\tilde H_j } \omega_{\tilde S_i }} {\rm Im } \left\{ \left[ (\varepsilon_{\tilde S_i } \varepsilon_{\tilde H_j}^*-k^2 ) C_A + M_{\tilde S_i } M_{\tilde H_j } C_B^{} \right]  { h(\varepsilon_{\tilde S_i }) \mp h(\varepsilon_{\tilde H_j }^*) \over \varepsilon_{\tilde S_i }-\varepsilon_{\tilde H_j }^* } \right. \nonumber \\&+&\left.\left[ (\varepsilon_{\tilde S_i } \varepsilon_{\tilde H_j}+k^2 ) C_A +  M_{\tilde S_i } M_{\tilde H_j } C_B^{} \right]  { h(\varepsilon_{\tilde S_i }) \mp h(\varepsilon_{\tilde H_j }) \over \varepsilon_{\tilde S_i }+\varepsilon_{\tilde H_j } }\right\}
\end{eqnarray}
with $C_A = 2 |\lambda_{ i j3 } \lambda_{i3j}|  v^2 $ and $ C_B = 2 |\lambda_{ i j3 } \lambda_{i3j}| \cos[ \arg(\lambda_{ij3})+ \arg(\lambda_{i3j} ) -\arg (\lambda_{3jj})] v^2 \sin 2 \beta$. It is straightforward to obtain the corresponding source term mediated by the new gaugino by making the following replacements:  $\lambda_{ij3} \to g^\prime \tan^{-1}\beta   Q_d  $, $\lambda_{i3j} \to g^\prime \tan\beta  Q_u  $, $\omega_{\tilde S_i } \to \omega_{\tilde B^\prime}$ and $M_{\tilde S_ i } \to M_{\tilde B^\prime}$. We assume no net density of gauginos and singlinos, thereby setting $\mu_{\tilde S_i } =\mu_{\tilde B^\prime}=0 $ and giving 
\begin{eqnarray}
S^{\rm CP }_{\tilde H } = - \Gamma_{h } \mu_{\tilde H }
\end{eqnarray}
where $\Gamma_{h } = \Gamma_{\tilde H}^- - \Gamma_{\tilde H}^+ $. 

We now derive the Boltzmann equations. We assume there are approximate chemical equilibriums between the SM particles and their superpartners, as well as between different members of left-handed fermion doublets. In this case, one obtains transport equations for densities associated with different members of supermultiplet. Since all light quarks are mainly produced by strong sphaleron processes and all quarks have similar diffusion constants, baryon number conservation on time scales shorter that the inverse electroweak sphaleron rate implies the approximate constraints $q_{1L} =q_{2L}=-2u_R=-2d_R=-2s_R=-2c_R=-2b_R\equiv -2b =  2( Q+T)$.  We define the number density for Higgs supermultiplet as
\begin{eqnarray}
H\equiv \sum_{i =1}^3 \left( n_{H_{u i}^+} + n_{H^0_{ui}} - n_{H_{di}^- } - n_{H_{di }^0} + n_{\tilde H_{ui}^+ } - n_{\tilde H_{d i }^-} + n_{\tilde H_{u i}^0} - n_{\tilde H_{di}^0 } \right) \; , 
\end{eqnarray}
The transport equation of the Higgs supermultiplet number density can  be written as
\begin{eqnarray}
\partial^\mu H_\mu = -\Gamma_h {H\over k_H } - \Gamma_Y \left( { Q \over k_Q } + {H \over k_H } -{  T \over k_Y }\right)  + S_{\tilde B^\prime }^{\rm CPV} + \sum_i S_{\tilde S_i }^{\rm CPV}
\end{eqnarray}
where $\partial^\mu =v_w {d \over d \bar z} -D_a {d^2 \over d \bar z^2}$ in the planar bubble wall approximation with $D_a$ the diffusion constant.  $n_i$ and $k_i$ is the number density and the statistical factor of particle $``i"$. 
The coefficient $\Gamma_{Y}$  denotes the interaction rate arising from top quark, which can be written as $\Gamma_Y=6|y_t|^2 I_F(m_{\tilde t_L}, m_{\tilde t_R}, m_h)$.  We refer the reader to \cite{Cirigliano:2006wh} for the general form of $I_F$.
$\Gamma_{h}$ denote the CP-conserving scattering rates of  Higgsinos with the background Higgs field within the bubble.

The CP-asymmetry Higgs supermultiplet number density produced by CP-violating interactions at the wall of the expanding bubbles may be transported into a net density of left handed fermions via the inelastic scattering.   By taking appropriate linear combinations of transport equations of the Higgs, the third generation left-handed quark doublet and right-handed top quark supermultiplets, we only need to solve a single equation for $H$.  Lastly, the BAU is given by
\begin{eqnarray}
n_B= -{3 \Gamma_{ws} \over 2 D_Q \lambda_{+} } \int^{-L_w/2}_{-\infty} dz n_L(z) e^{-\lambda_{-} z }
\end{eqnarray}
with  $\lambda_{\pm} ={1 \over 2 D_Q} (v_w\pm \sqrt{v_w^2 +4D_Q R})$,  where $R\sim 2\times 10^{-3}~{\rm GeV}$ is the { inverse} washout rate for the electroweak sphaleron transitions. 

\begin{table}[h]
\centering
\begin{tabular}{cc|cc |cc|cc}
\hline \hline 
$T$ & $100~{\rm GeV}$ & $\Delta \beta$ & $0.01$  & $D_Q$ & $6/T$ &$ \Gamma_{\tilde t }$ & $0.25T$ \\
$v(T)$ & 125~${\rm GeV}$ & $v_w$ & 0.2 & $D_H$ &$100/T$  & $\Gamma_{\tilde Z^\prime}$ & $0.030T$ \\
$L_w $ & $0.25/T$ &$\tan \beta $ &$10$ &$A_t$ & 300 & $\Gamma_{\tilde H_3 }$ & $0.025T$ \\
\hline \hline
\end{tabular}
\caption{ Input parameters at the benchmark point.  }\label{aaa}
\end{table}

The computation of $n_B/s$ relies upon many other numerical inputs; our choices are listed in Table. I. The bubble wall velocity $v_w$, thickness $L_w$, profile parameters $\Delta \beta$ and $v(T)$ describe the dynamics of the expanding bubbles during the EWPT, at the temperature $T$. We take the Higgs profile to be
\begin{eqnarray}
v(z) &\simeq&  {1 \over 2 } v(T) \left\{ 1 + {\rm tanh}\left( 2 \alpha {z \over L_w}\right)\right\} \; ,  \\
\beta(z) &\simeq& \beta_0 (T) - {1 \over 2 } \delta \beta \left\{1 -  {\rm tanh}\left( 2 \alpha {z \over L_w}\right)\right\} \; ,
\end{eqnarray}
following Ref.~\cite{Moreno:1998bq,Carena:2000id,Carena:1997gx}.  The sphaleron rates are $\Gamma_{ws} =6 \kappa \alpha_s^4 T$ and $\Gamma_{ss}=6 \kappa^\prime \alpha_s^4 T {8 \over 3 } $,  where $\kappa_{ws} = 22\pm 2$ and $\kappa_{ss} ={\cal O }(1)$.  
%%%%%%%%%%%%%%%%%%%%%%%%%%%%%%%%%%%%%%%%%%%%%%%%%%%%%%%%%%%%%%%%%%%%%%
\begin{figure}[t]
\begin{center}
\includegraphics[width=7.0cm,height=5cm,angle=0]{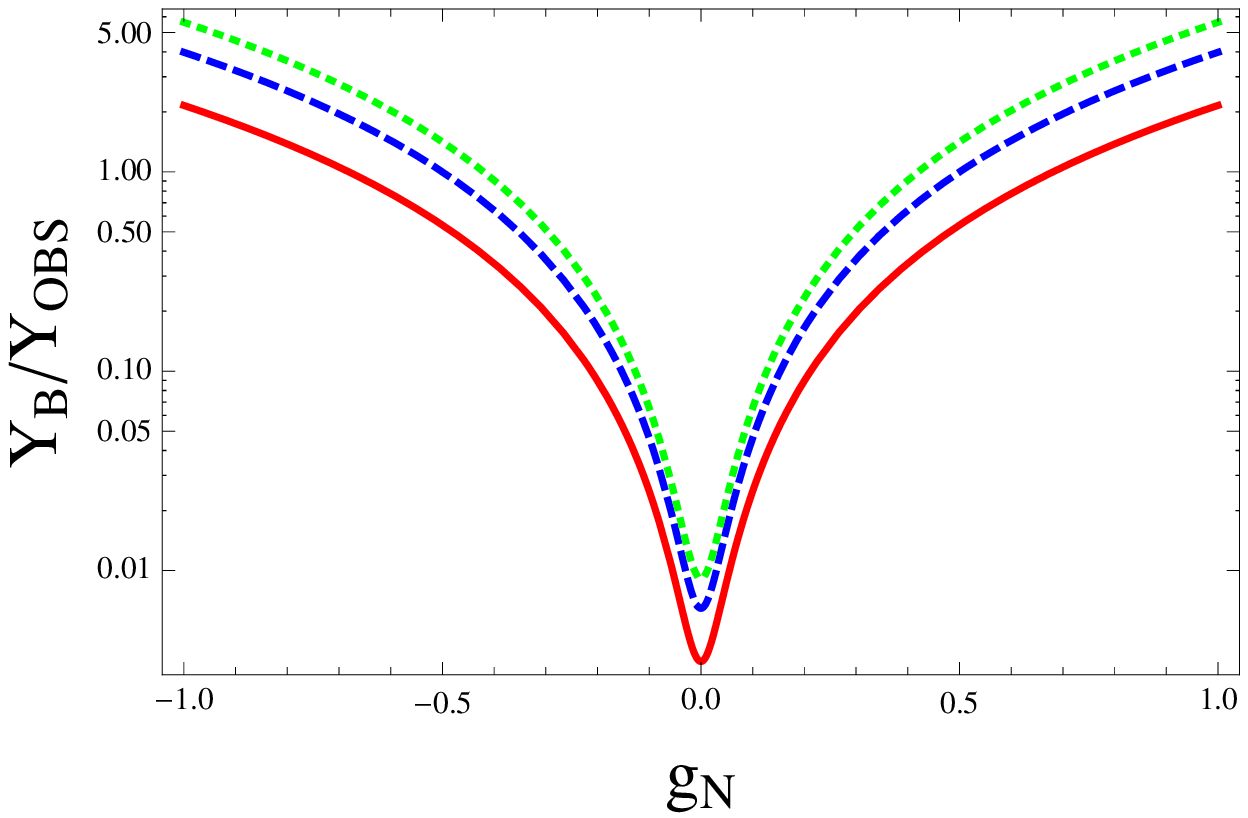}
\hspace{1.0cm}
\includegraphics[width=7.0cm,height=5cm,angle=0]{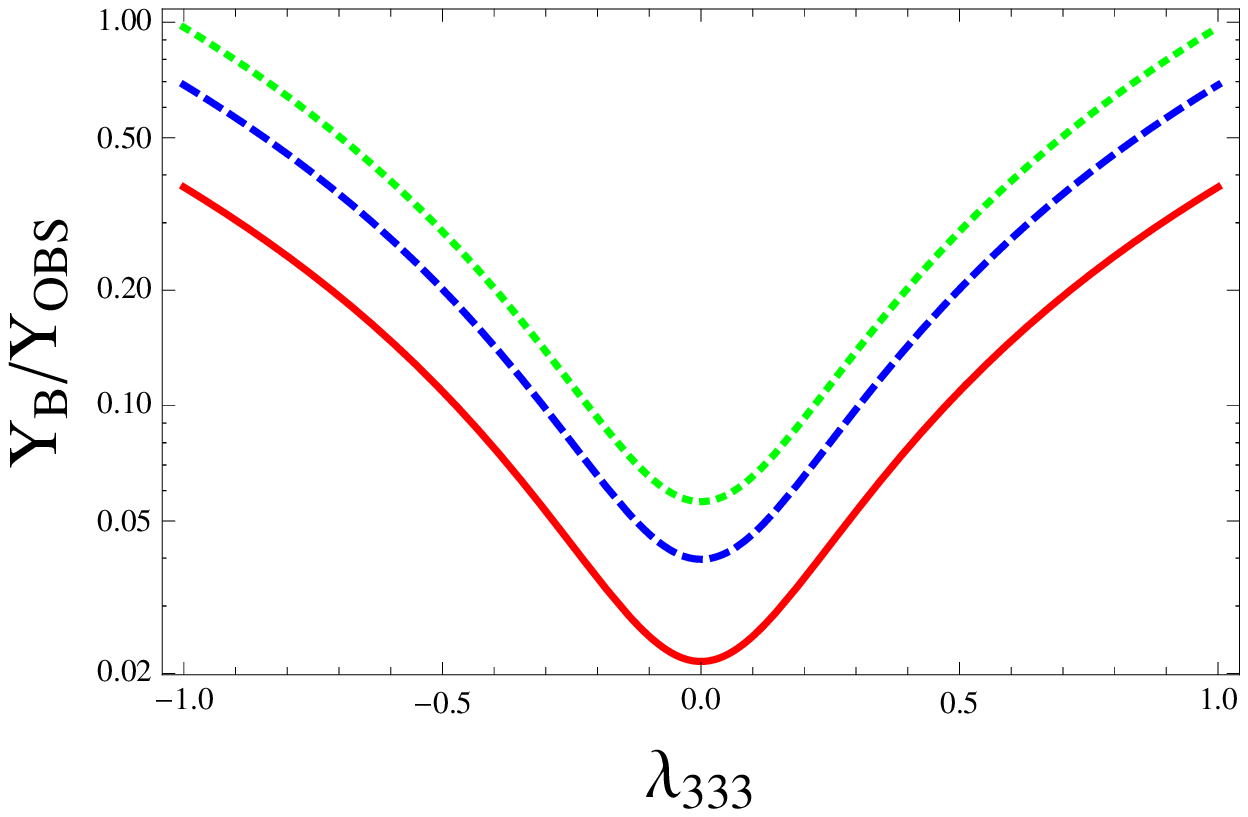}
\end{center}
\caption { $Y_B/Y_{\rm obs}$ as the function of $g_N$ (left panel) and $\lambda_{333}$ (right panel) by setting $\phi= {\pi/8}$ (solid),${\pi/4}  $  (dashed) and $\pi/2$ (dotted) respectively. We  have set $M_{\tilde Z '} $ =400~${\rm GeV}$ and $M_{\tilde H_3}$ = 350~${\rm GeV}$.}\label{figlines}
\end{figure}
%%%%%%%%%%%%%%%%%%%%%%%%%%%%%%%%%%%%%%%%%%%%%%%%%%%%%%%%%%%%%%%%%%%%%%

%%%%%%%%%%%%%%%%%%%%%%%%%%%%%%%%%%%%%%%%%%%%%%%%%%%%%%%%%%%%%%%%%%%%%%
%\begin{figure}[h]
%\begin{center}
%\includegraphics[width=4.5cm,height=3.5cm,angle=0]{contgN.eps}
%\hspace{0.5cm}
%\includegraphics[width=4.5cm,height=3.5cm,angle=0]{contl3.eps}
%\hspace{0.5cm}
%\includegraphics[width=4.5cm,height=3.5cm,angle=0]{contgnlam.eps}
%\\
%(a) \hspace{5cm} (b) \hspace{5cm} (c)
%\end{center}
%\caption{Contours of $Y_B/Y_{\rm obs}$ in the $\phi-g_N$ (a), $\phi-\lambda_{333}$ (b) and $\lambda_{333}-g_N$ (c) planes. We have  set $M_{\tilde Z '} $ =400~${\rm GeV}$ and $M_{\tilde H_3}$ = 350~${\rm GeV}$. }\label{contourA}\end{figure}
%%%%%%%%%%%%%%%%%%%%%%%%%%%%%%%%%%%%%%%%%%%%%%%%%%%%%%%%%%%%%%%%%%%%%%

%%%%%%%%%%%%%%%%%%%%%%%%%%%
\begin{figure}[h!]
%\centering
\subfloat[]{
\includegraphics[width=0.32\textwidth]{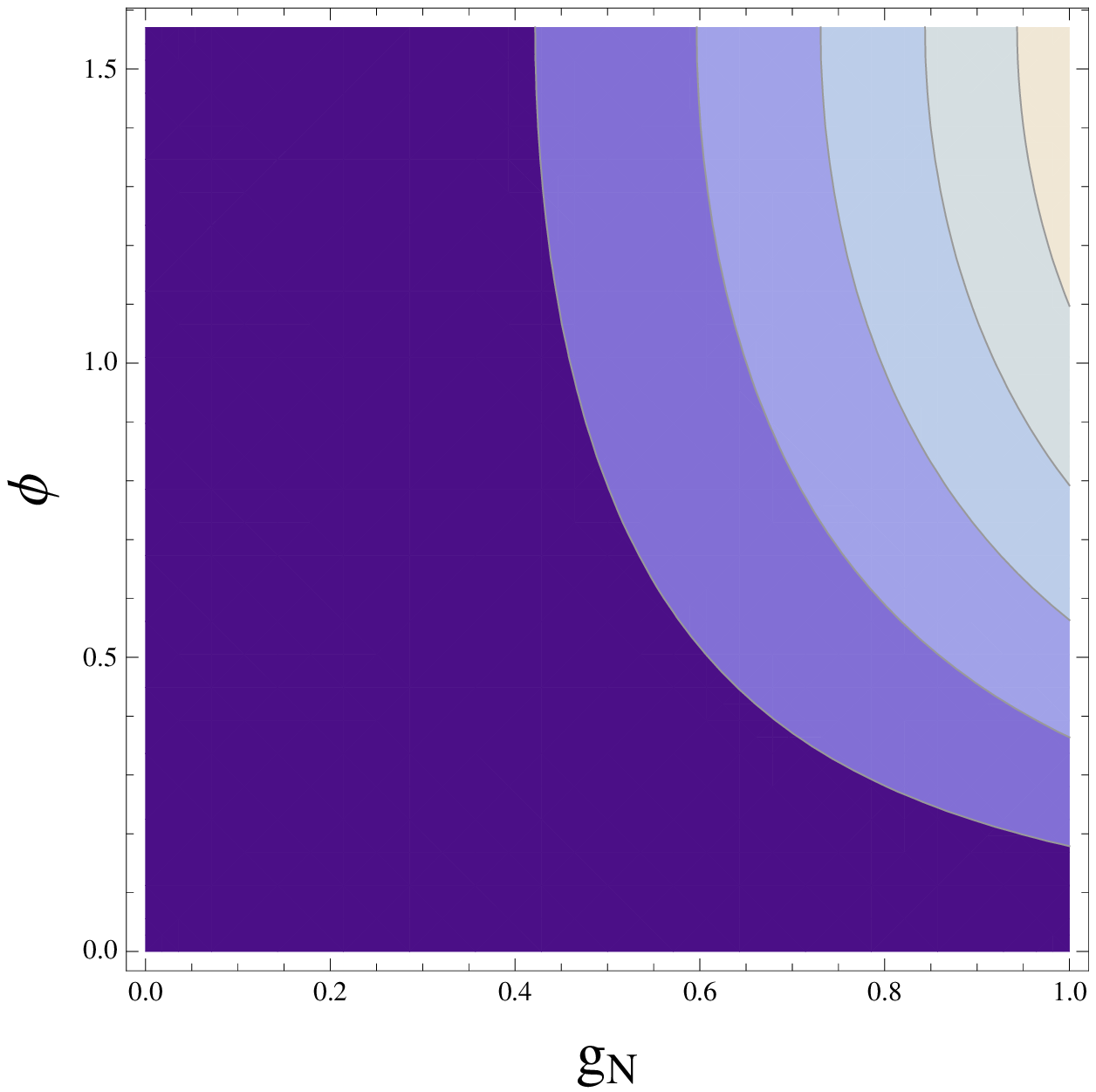}}
~~\subfloat[]{
\includegraphics[width=0.32\textwidth]{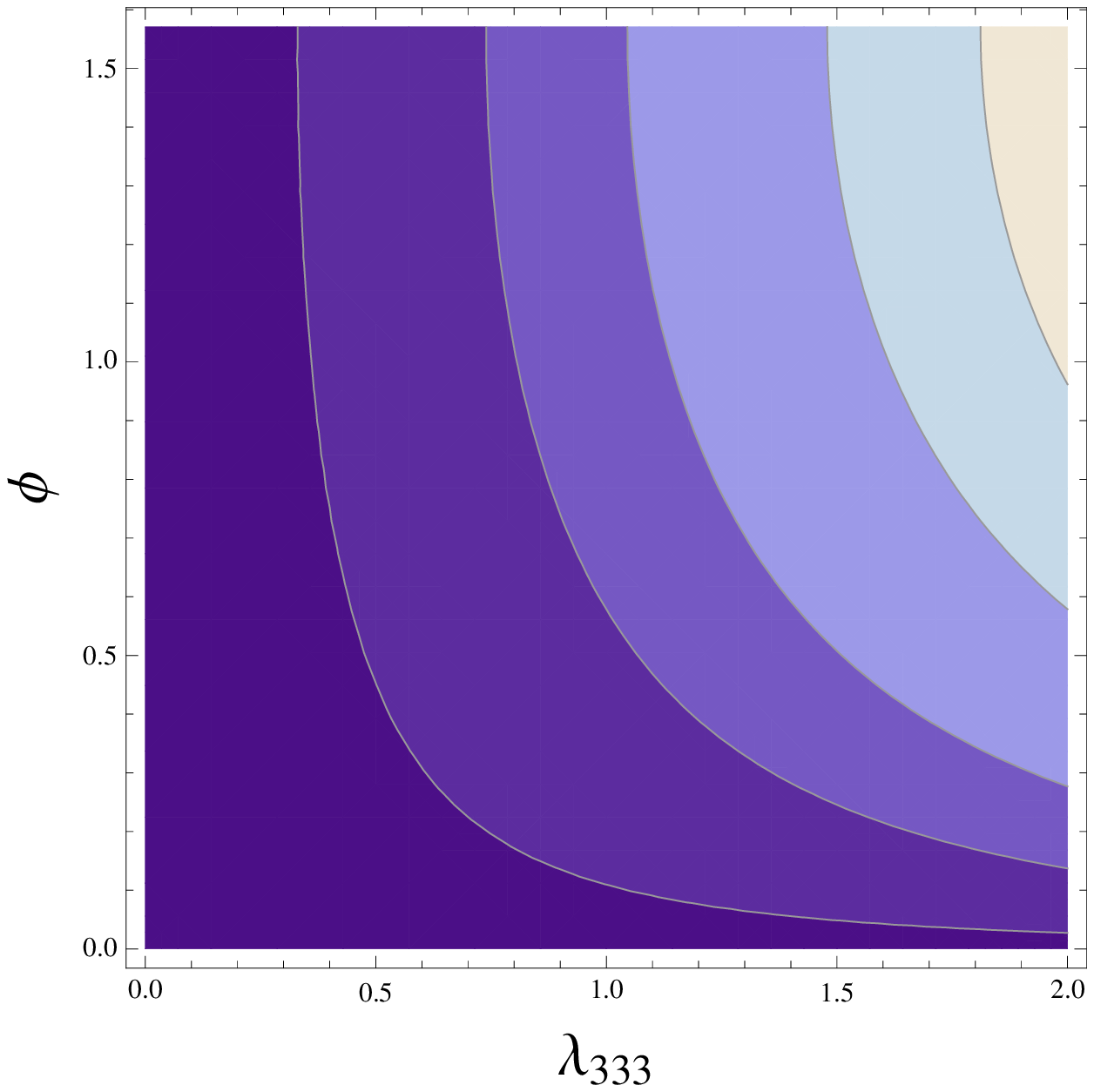}}
~~\subfloat[]{
\includegraphics[width=0.32\textwidth]{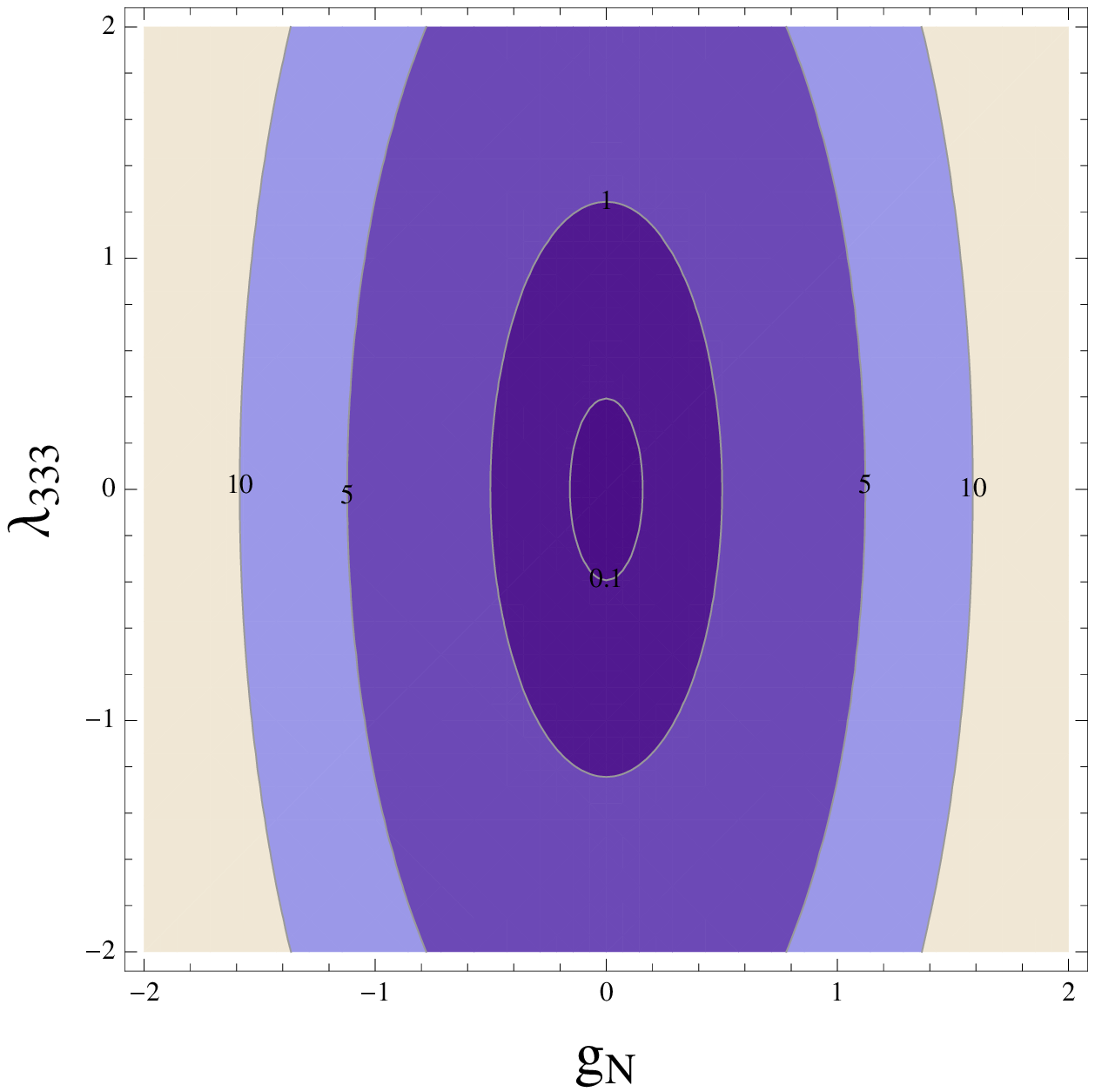}}
\caption{Contours of $Y_B/Y_{\rm obs}$ in the $\phi-g_N$ (a), $\phi-\lambda_{333}$ (b) and $\lambda_{333}-g_N$ (c) planes. We have  set $M_{\tilde Z '} $ =400~${\rm GeV}$ and $M_{\tilde H_3}$ = 350~${\rm GeV}$.} \label{contourA}
\end{figure}
%%%%%%%%%%%%%%%%%%%%%%%%%%%

\begin{figure}[t]
\begin{center}
\includegraphics[width=7.0cm,height=5cm,angle=0]{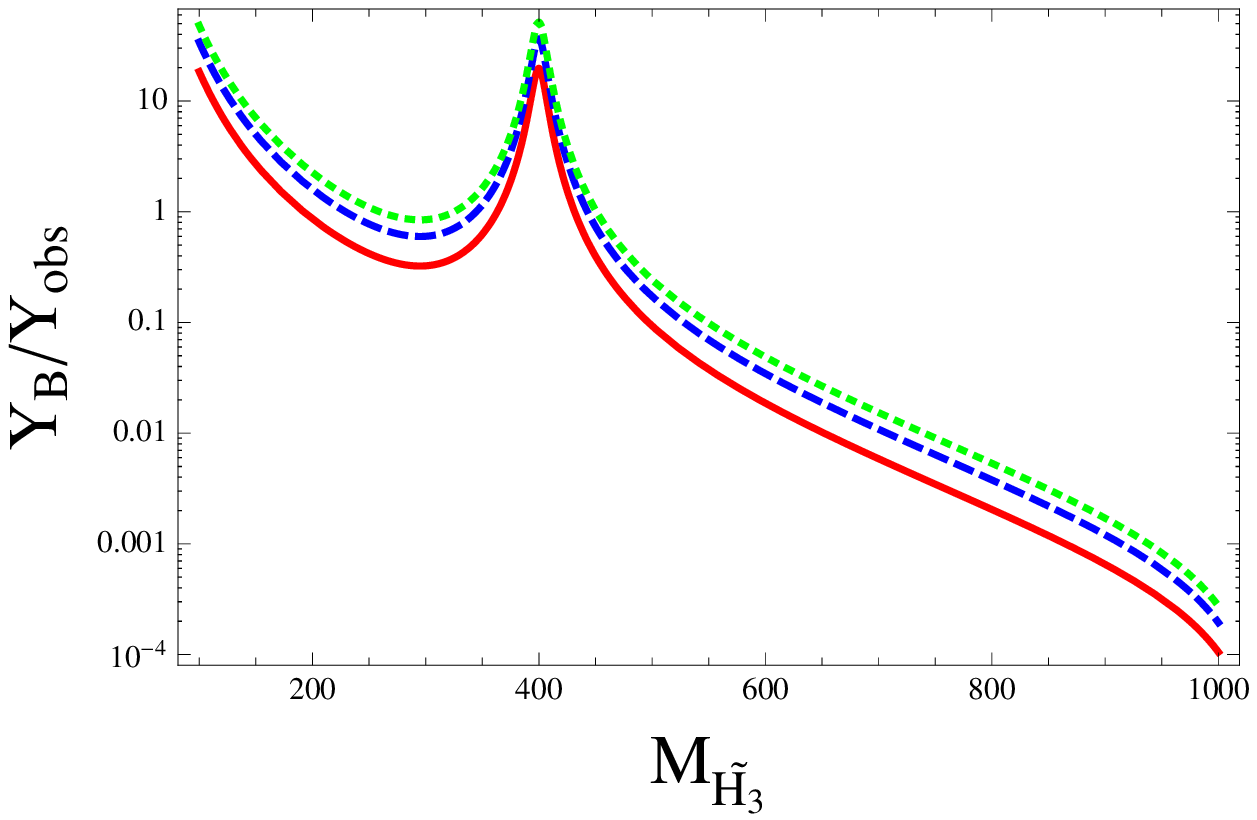}
\hspace{1.0cm}
\includegraphics[width=7.0cm,height=5cm,angle=0]{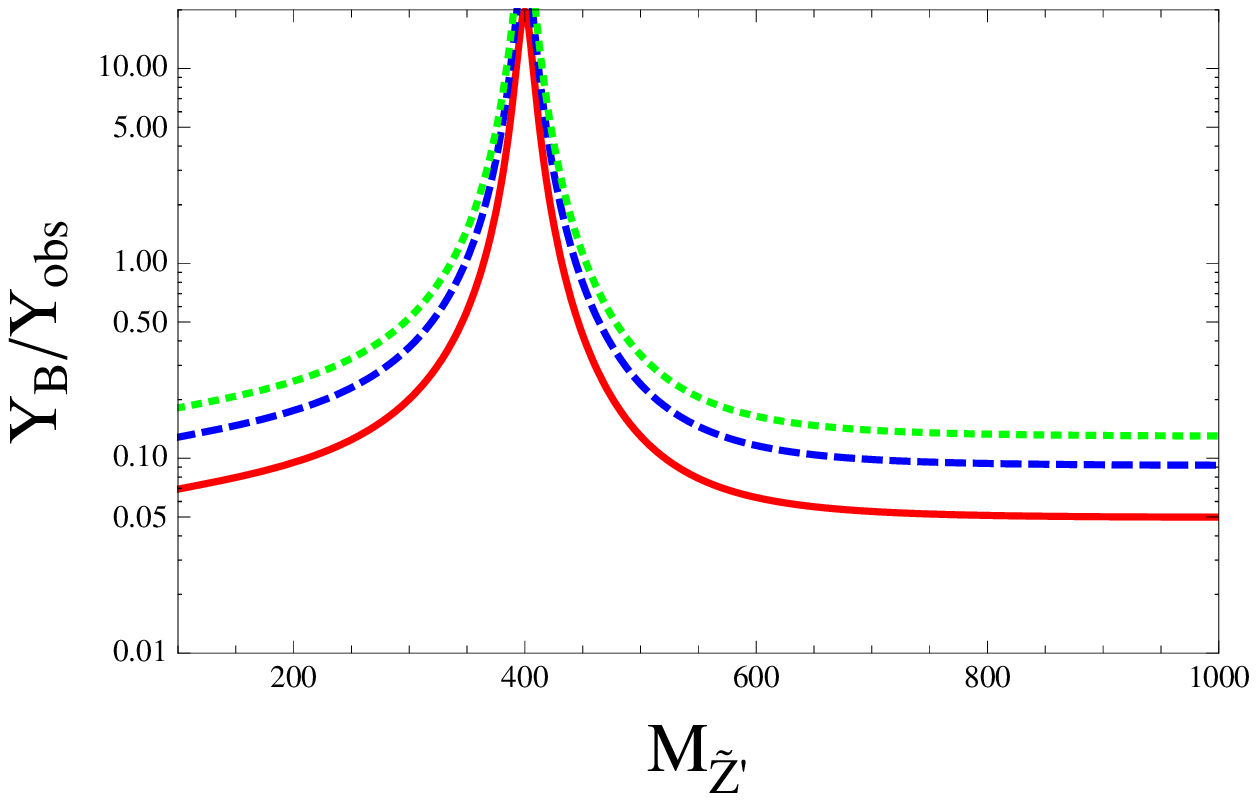}
\end{center}
\caption{ $Y_N/Y_{\rm obs}$ as the function of $M_{\tilde H_3}$  (left panel)and $ M_{\tilde Z^\prime}$, where the solid, dashed and dotted lines correspond to $\phi=\pi/8,~\pi/4$ and $\pi /2$ respectively.}\label{mass}\end{figure}

For simplification, we assume only  $\lambda_{333}$ contains a complex phase. We show in the left (right) panel of  Fig. \ref{figlines}, $Y_B/Y_{\rm obs}$, where $Y_{\rm obs}$ is the observed value given in eq(\ref{expp}), as the function of $g_N $ $(\lambda_{333})$, by setting $\lambda_{333} ~(g_N)=0.1$, other initial inputs are given in Table. \ref{aaa}. The solid, dashed and dotted lines correspond to $\phi =\pi/8,\pi/4, ~{\rm and}~\pi/2$, respectively.  Obviously the observed baryon asymmetry can be obtained by these two CP-violating sources separately.  To study the relative contribution of $\tilde Z^\prime$ and $\tilde S_3$ to the baryon asymmetry, we show contours of $Y_B/Y_{\rm obs}$ in the $\phi-g_N$ plane( Fig. \ref{contourA} (a) ) and in the $\phi-\lambda_{333}$ plane (Fig. \ref{contourA} (b)).  Contours from left to right correspond to $Y_B/Y_{\rm obs} =1,~2,~3,~4,~5$ (Fig. \ref{contourA} (a)) and $Y_B/Y_{\rm obs} =0.1,~0.5,~1,~2,~3$ (Fig. \ref{contourA} (b)) respectively. We show contours of  $Y_B/Y_{\rm obs}$ in the $g_N-\lambda_{333}$ plane by assuming $\phi=\pi/4$ in the Fig. \ref{contourA} (c).  Notice that the new gaugino induced CP-violating source is more effective to give  rise to a sizable baryon asymmetry. This is because we set a narrower mass splitting between $M_{\tilde H_3}$ and $M_{\tilde Z' }$ than that between $M_{\tilde H_3}$ and $M_{\tilde S_3}$ when carrying out numerical calculation.   We show in the left (right) panel of Fig. \ref{mass} $Y_B/Y_{\rm obs}$ as the function of $M_{\tilde H_3^0}$ ($M_{\tilde Z^\prime}$) by assuming $M_{\tilde Z^\prime} (M_{\tilde H_3^0})=400~{\rm GeV}$ and $g_N=\lambda_{333}=0.5$.  The solid dashed and dotted lines correspond to $\phi=\pi/8,\pi/4,\pi/2$ respectively. It is obvious that there is a resonant enhancement to the production of the  baryon asymmetry when $M_{\tilde H_3^{}}  = M_{\tilde Z^\prime}$.   

Finally let us consider  constraints on the CP phases of the neutralino mass matrix from the non-observation of the electric dipole moments for neutrons and the electron.  These CP-violating phases may contribute to EDMs via the $H^+W^-$ or $W^+W^-$ mediated Bar-Zee graphs.   It was observed in Ref.~\cite{Li:2008ez}  that CP violation in the bino-Higgsino sector of the MSSM can account for successful baryogenesis without inducing EDMs. This observation weaken the correlation between the electroweak baryogenesis and EDMs.  It was found \cite{Cheung:2011wn,Cheung:2012pg} that  the maximal CP phase $\phi$ is still compatible with the current EDM constraints in the NMSSM. The same argument can be applied to our model since we only focus on singlinos and the new gaugino induced CP-violating source terms. We leave the  systematic study of constraints of EDMs in  the ${\rm E_6SSM}$ to another project.

%%%%%%%%%%%%%%%%%%%%%%%%%%%%%%%%%%%%%%%%%%%%%%%%%%%%%%%%%%%%%%%%%%%%%%
%%%%%%%%%%%%%%%%%%%%%%%%%%%%%%%%%%%%%%%%%%%%%%%%%%%%%%%%%%%%%%%%%%%%%%

\section{Conclusion}

MSSM has difficulty in explaining both electroweak baryogenesis and $125$ GeV Higgs.  Possible extensions to the MSSM accounting these two problems were  well studied recently. In this paper, we studied electroweak baryogenesis in the $E_6$ inspired supersymmetric standard model, which contains at least two more CP-violating source terms in the neutralino sector compared with the MSSM case.  New CP-violationgs source terms as well as transport equations  of the Higgs supermultiplet were calculated analytically and numerically. Our results show that  CP-violating sources from singlinos and the new gaugino can give rise to a  successful electroweak baryogenesis respectively.   It should be mentioned that we only studied the adequate condition for a successful baryogenesis in the ${\rm E_6SSM}$. A systematic study of the EDMs constraint to the ${\rm E_6SSM}$, which is important and necessary but beyond the reach of this paper, will be shown in an another paper. 

\begin{acknowledgments}
This work was supported in part by DOE Grant DE-SC0011095.
\end{acknowledgments}

%\begin{eqnarray}
%\left(  \matrix{\times & 0 & \times & \times & 0&0&0&0&0&0&0&0 \cr
%0 & \times & \times & \times & 0&0&0&0&0&0&0&0\cr
%\times & \times & 0 & \times & \times &\times&0&\times&\times&0&\times&\times\cr
%\times&\times&\times&0&\times&\times&\times&0&\times&\times&0&\times\cr
%0&0&\times&\times&0&\times&\times&\times&0&\times&\times&0\cr 
%0&0&\times&\times&\times&\times&0&0&0&0&0&0\cr
%0&0&0&\times&\times&0&0&\times&\times&0&\times&\times\cr
%0&0&\times&0&\times&0&\times&0&\times&\times&0&\times\cr
%0&0&\times&\times&0&0&\times&\times&0&\times&\times&0\cr
%0&0&0&\times&\times&0&0&\times&\times&0&\times&\times\cr
%0&0&\times&0&\times&0&\times&0&\times&\times&0&\times\cr 
%0&0&\times&\times&0&0&\times&\times&0&\times&\times&0} \right)
%\end{eqnarray}

\end{document}